\documentclass[doublecol,figures]{epl2}

\usepackage{graphicx,amssymb,amsmath,setspace,stmaryrd,bm}

\title{Confinement-induced enhancement of diffusiophoretic forces on
self-propellers}

\author{
M. N. Popescu\inst{1,2,3}\thanks{E-mail: \email{popescu@mf.mpg.de}}
\and S. Dietrich\inst{1,2}\thanks{E-mail: \email{dietrich@mf.mpg.de}}
\and G. Oshanin\inst{4,1,2}\thanks{E-mail: \email{oshanin@lptl.jussieu.fr}}
}

\shortauthor{M. N. Popescu, S. Dietrich, and G. Oshanin}

\institute{
\inst{1} Max-Planck-Institut f\"ur Metallforschung - Heisenbergstr. 3,
70569 Stuttgart, Germany\\
\inst{2} Institut f\"ur Theoretische und Angewandte Physik,
Universit\"at Stuttgart - Pfaffenwaldring 57, 70569 Stuttgart,
Germany\\
\inst{3} Ian Wark Research Institute, University of South Australia,
Mawson Lakes (Adelaide), SA 5095, Australia\\
\inst{4} Laboratoire de Physique Th{\'e}orique de la Mati{\`e}re
Condens{\'e}e, Universit{\'e} Pierre et Marie Curie (Paris 6) -
4 Place Jussieu, 75252 Paris, France
}


\abstract{
We study the effect of spatial confinement on the strength of
propulsive diffusiophoretic forces acting on a particle that
generates density gradients by exploiting the chemical free energy
of its environment. Using a recently proposed simple model of a
self-propelling device driven by chemical reactions taking place
on some parts of its surface, we demonstrate that the force
significantly increases in the presence of confining walls.
We also show that such effects become even more pronounced in
two-dimensional systems.
}

\begin{document}

\maketitle

\label{intro} Last years have witnessed a growing technological,
experimental, and theoretical interest in scaling standard machinery
down to micro- and nano-scales, such as producing pumps, motors, or
sieves needed for the development of ``lab on a chip'' devices. For
applications in, e.g., drug-delivery systems or micromechanics
\cite{Paxton_2005,Paxton_2006}, one of the most challenging problems
at this stage is to develop ways to enable small-scale objects to
perform autonomous, controlled motion. Exploiting a variety of
physical and chemical processes, a number of minimalistic, ``proof
of principle" proposals of such micro-engines have been put forward
for both translational and rotational motion. Although the research
in this area is still in its early stages, several of these proposals
have already been tested experimentally (see, e.g.,
Refs.\cite{Whitesides_2002,Sen_2005}; for a recent review, see, e.g.,
Ref.\cite{Paxton_2006}).

Among the key issues in the development of such engines are their
``fuelling'' and the tuning of the driving force and torque.
Inspired by biological nano-engines that are using catalytic
reactions to extract energy from the environment (e.g., kinesin moving
along microtubules is a molecular motor that uses hydrolization
of ATP as the energy source), Whitesides and co-workers proposed a
design of self-propelling devices based on an asymmetric decoration
of the surface of small objects by catalytic particles promoting a
chemical reaction in the surrounding medium \cite{Whitesides_2002}.
The asymmetry in the catalyst placement leads to an asymmetry in
the distribution of the reaction products, eventually also in the
distribution of reactants, and this asymmetric distribution can
provide motility through a variety of mechanisms, such as surface
tension gradients, cyclic adsorption and desorption, and
diffusiophoresis \cite{Paxton_2006,Kapral_2007,Golestanian_2007};
moreover, such a design removes the need of a ``fuel reservoir''
within the device because the reactants are provided by the
environment.

Following the idea of the asymmetric surface distribution of catalysts,
Golestanian \textit{et al} \cite{Golestanian_2005} have recently
studied a model system in which the autonomous motion emerges as
a result of self-created diffusiophoretic gradients. Specifically,
they considered a spherical particle of radius $R$ with a single,
point-like catalytic site fixed on its surface, which is immersed
in the three-dimensional (\textit{3d}) bulk of a reactive solvent.
The catalytic site was deemed to promote within the solute a
chemical reaction that creates product particles of a size much
smaller than $R$. The spatial distribution of the emerging
(diffusive) product particles is asymmetric and thus results in
an effective diffusiophoretic force acting on the particle.
Several aspects of the resulting motion have been discussed, such
as the ensuing transient time-dependent behavior, effects due to
density and reaction-rate fluctuations, as well as the influence of
rotational diffusion on the directionality \cite{Golestanian_2005}.
A first experimental realization of this model, using
platinum coated polystyrene spheres, has been recently reported.
\cite{Howse_2007}

For most of the applications in biological systems or 'lab on a
chip'-type devices one has to deal with a complicated internal
structure of the system, e.g., networks of narrow channels or pores
and various impenetrable impurities. Thus the spatial confinement is
a relevant feature which may even lead to a quasi two- or
one-dimensional behavior so that assuming the presence of an
unconfined \textit{3d} bulk reactive solvent, as in
Refs. \cite{Paxton_2005,Paxton_2006,Golestanian_2005}, may break
down. Moreover, studies of bacteria or cell motion on surfaces such
as nutrient substrates represent examples of two-dimensional systems.
A study of two-dimensional self-propelled particles with fluctuations
in speed and direction of motion, aiming at describing such
experimental systems has been recently reported. \cite{Peruani_2007}
Intuitively, one expects that the spatial confinement or the reduced
dimensionality may influence the resulting motion of self-propelling
objects like the ones discussed above, but {\it a priori} it is not
clear if this will lead to a decrease or an increase in the value of
the diffusiophoretic force acting on the particle, and if such changes
are of relevant magnitude.

Based on the model used in Ref. \cite{Golestanian_2005}, here we
address the effects of spatial confinement on the diffusiophoretic
force exerted on a self-propelling particle. For a catalytic site
with a uniform, i.e., time-independent reaction rate we show
\textbf{(1)} that in \textit{3d} the presence of a confining wall
for the reaction product leads to a significant increase of the
force, even in the case of the simplest possible confining geometry
given by a concentric wall with the same shape as the particle, and
\textbf{(2)} that a reduced dimensionality alone, i.e., a
two-dimensional (\textit{2d}) system, also strengthens the force
acting on the particle.

\begin{figure}[!htb]
\onefigure[width=.9\linewidth]{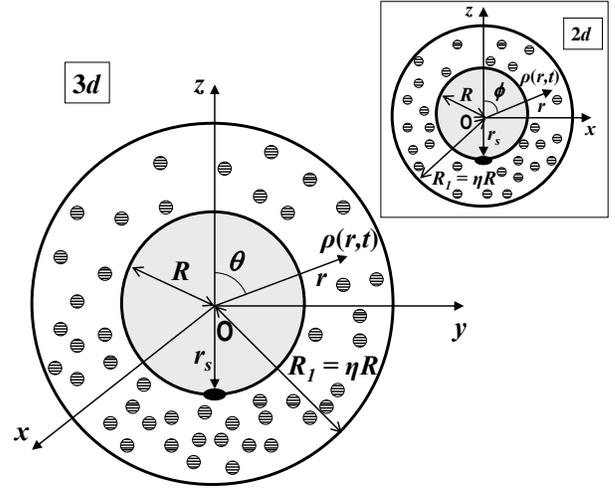}
\caption{
An impermeable, spherical (disk-like in \textit{2d}) particle of
radius $R$ with a point-like catalytic site at ${\vect{r}}_s$
(depicted as a black dot) on its surface (perimeter in \textit{2d}),
enclosed by a concentric, impermeable, spherical (circular
in \textit{2d}) wall of radius $R_1 = \eta R$ ($\eta > 1)$. The
top inset shows the notations for the \textit{2d} case. The reaction
products, indicated by small hatched circles, exhibit a gradient
of their number density $\rho(\vect{r},t)$ (per volume in \textit{3d}
and per area in \textit{2d}). This setup can also be thought of as an
elementary micropump creating a microcurrent into a confined
environment.
}
\label{fig1}
\end{figure}

\label{Model} The system we consider is shown in Fig.~\ref{fig1}.
It consists of an impermeable, spherical particle (disk in
\textit{2d}) of radius $R$ with a point-like catalytic
site (black dot in Fig.~\ref{fig1}) on its surface (perimeter
in \textit{2d}), which promotes a chemical conversion of a
surrounding solvent into product particles (small hatched circles
in Fig.~\ref{fig1}). The particle and the surrounding solvent are
enclosed in a concentric, impermeable, spherical (circular
in \textit{2d}) shell of radius $R_1 = \eta R$ ($\eta > 1)$.
Note that in \textit{3d} the model discussed in
Ref.~\cite{Golestanian_2005} is recovered in the limit
$\eta \to \infty$.
The whole system is assumed to be \textit{rigidly fixed} at the
common center O, such that we do not have to account for
effects due to the motion through the surrounding medium (e.g.,
if the solvent is a liquid, hydrodynamic interactions may
also play a role), i.e., as a first step here we focus on the
statics of the self-propelling device.

Following Ref.~\cite{Golestanian_2005}, we assume that the
reaction at the catalytic site acts effectively only as a
point-like source of product molecules, which are diffusing in
the solvent with diffusion coefficient $D$; the amplitude of the
production rate of this source, located at
${\vect{r}}_s = - R \hat{\vect{z}}$ (where $\hat{\vect{z}}$ is
the unit vector of the $z$-axis), is denoted as $A(t)$. Accordingly,
the time evolution of the density $\rho(\vect{r},t)$ of product
molecules is governed by the diffusion equation
\begin{equation}
\label{diff_eq}
\partial_t \rho = D \nabla^2 \rho +
A(t) \delta^3(\vect{r}-{\vect{r}}_s),~~R < |\vect{r}| < R_1,
\end{equation}
subject to the initial condition (IC) of zero density of product
molecules and to the boundary conditions (BC) of zero normal current
on the surface of the particle and on the confining wall:
\begin{subequations}
\label{BC_IC}
\begin{eqnarray}
&&\rho(\vect{r},0) = 0,\\
&&\left.\left(
\frac{\partial \rho(\vect{r},t)}{\partial r}
\right)\right|_{|\vect{r}|= R, R_1} = 0.
\end{eqnarray}
\end{subequations}

As discussed in Refs.~\cite{Anderson_1989,Golestanian_2005}, an
asymmetric, non-uniform distribution of products $\rho(\vect{r},t)$
induces particle motion with a velocity proportional to the
first moment of the density distribution $\rho(\vect{r},t)$ on the
surface of the object. Since this is the result of a Stokes
force-like calculation, we operationally define for our system,
which is rigidly fixed, the net diffusiophoretic force acting on the
particle at time $t$ along the $z$-direction as also being
proportional to the first moment of $\rho(|\vect{r}|= R,t)$:
\begin{subequations}
\label{def_force}
\begin{eqnarray}
F_z(t) & = &- 2 \pi R^2 b_3 \int\limits_{0}^{\pi}
\upd \theta \,\sin\theta\, \cos\theta \,
\rho(R,\theta,t),~~\textrm{in } \textit{3}d,~~~~~~~~\label{def_force_3d}\\
F_z(t) & = &- R b_2 \int\limits_{0}^{2\pi}
\upd \phi \,\cos\phi\,\rho(R,\phi,t),~~\textrm{in } \textit{2}d,
\label{def_force_2d}
\end{eqnarray}
\end{subequations}
where $b_3$ and $b_2$ are some system-dependent constants with the
dimension of energy and of the order of the thermal energy $k_B T$
\cite{note1} (in the following, we set $b = b_3 = b_2$ for simplicity).
Knowledge of $F_z(t)$ as a function of $\eta < \infty$ for
confined \textit{3d} and \textit{2d} systems allows us to quantify
the effect of confinement on the propulsive force, and thus we proceed
by computing $F_z(t)$ for general values of $\eta$.

The force is known once the density $\rho(\vect{r},t)$ [i.e., the
solution of Eq.(\ref{diff_eq}) subject to the IC and BC conditions
in Eq. (\ref{BC_IC})] is determined. Using the Laplace transform and
the inversion theorem \cite{Carslaw_book}:
\begin{equation}
\bar f(p) \equiv {\cal L} [f]
=  \int\limits_0^\infty \upd  t \,e^{-p t} f(t) \label{LT}
\end{equation}
and
\begin{equation}
f(t)\equiv {\cal L}^{-1}[\bar f(p)]:=
\frac{1}{2\pi i}
\int\limits_{\gamma-i \infty}^{\gamma + i \infty}
\upd p \,e^{p t}\bar f(p),\label{ILT}
\end{equation}
respectively, where $\bar f(p)$ is assumed to be well-defined for
$p \in \mathbb{R}^+$ and $\gamma \in \mathbb{R}$ is sufficiently
large such that all singularities of $\bar f$ lie to the left of
the integration path, one finds that the Laplace-transformed
density is given by $\bar \rho(\vect{r},q)
= - \bar a(q) \bar G(\vect{r},q)$, where $q = \sqrt{p/D} > 0$,
$\bar a(q) = \bar A(q)/D$, and $\bar G(\vect{r},q)$ is the Green's
function for the Helmholtz operator $\nabla^2 - q^2$ with BCs of
vanishing normal derivative at $|\vect{r}|= R, R_1$. Decomposing
$\bar G(\vect{r},q)$ as $\bar G(\vect{r},q) = G_s(\vect{r},q) +
g(\vect{r},q)$, where the singular part $G_s$ is the free space
Green's function for the Helmholtz operator (which is known in
any spatial dimension $d$, see, e.g., Ref. \cite{Hassani_book}),
the initial problem has been reduced to that of finding the
solution $g$ of the homogeneous Helmholtz equation subject to the
boundary conditions
\begin{equation}
\label{BC_for_g}
\left.\left(
\frac{\partial g(\vect{r},q)}{\partial r}
\right)\right|_{|\vect{r}|= R, R_1}
= -\left.\left(
\frac{\partial G_s}{\partial r}
\right)\right|_{|\vect{r}|= R, R_1}\,.
\end{equation}
While the main steps in the calculation of $g$ are the same in
\textit{3d} and in \textit{2d}, the details do depend on the
dimensionality; for clarity, we will consider the two cases
separately.\\

\noindent{\it The three-dimensional case}\\
\indent In \textit{3d} the singular part of the Green's function is
given by \cite{Hassani_book}
\begin{equation}
\label{Gs_3d}
G_s(\vect{r},q;{\vect{r}}_s) = -\frac{q}{4\pi}
\frac{e^{-q |\vect{r}-{\vect{r}}_s|}}{q |\vect{r}-{\vect{r}}_s|},
\end{equation}
while the regular part (i.e., the general solution of the homogeneous
Helmholtz equation) can be written as
\begin{eqnarray}
\label{g_3d}
g(r,\theta,q;{\vect{r}}_s,\eta)
&=& \sum\limits_{\ell \geq 0}
[\alpha_\ell(q;{\vect{r}}_s,\eta) i_{\ell+1/2}(q r)\\\nonumber
&+& \beta_\ell(q;{\vect{r}}_s,\eta) k_{\ell+1/2}(q r)] P_\ell(\cos\theta),
\end{eqnarray}
where $i_{\ell+1/2}(z) = \sqrt{\pi/(2 z)} I_{\ell+1/2}(z)$ and
$k_{\ell+1/2}(z) = \sqrt{\pi/(2 z)} K_{\ell+1/2}(z)$ are the
modified spherical Bessel functions of the first and third kind,
respectively \cite{Abramowitz_book}, $P_\ell$ is the Legendre
polynomial of degree $\ell$, while the functional coefficients
$\alpha_\ell(q;{\vect{r}}_s,\eta)$ and $\beta_\ell(q;{\vect{r}}_s,\eta)$
will be fixed to fulfill the boundary conditions.

Using one of the addition theorems for the Bessel
functions \cite{Carslaw_book}, and noting that the angle between
$\vect{r}$ and ${\vect{r}}_s$ is $\pi - \theta$ (see Fig.~\ref{fig1}),
$G_s$ can be re-written as
\begin{equation}
\label{Gs_3d_expand}
G_s(\vect{r},q;{\vect{r}}_s) = - \sum\limits_{\ell \geq 0} c_\ell(q)
i_{\ell+1/2}(q r_<) k_{\ell+1/2}(q r_>) P_\ell(\cos\theta)
\end{equation}
where $c_\ell(q) = (-1)^\ell (2\ell+1) q / (2\pi^2)$,
$r_< = \min(r,r_s)$, and $r_> = \max(r,r_s)$. By combining
Eqs. (\ref{g_3d}, \ref{Gs_3d_expand}) and the boundary conditions
[Eq. (\ref{BC_for_g})], by noting that $r_s = R_+$ (i.e., the
source is \textit{on} the spherical surface at $r = R$ so that
$r_s - R_+ = 0^+$ ), and by re-writting the functional coefficients as
$\alpha_\ell \equiv c_\ell i_{\ell+1/2}(q R) \bar \alpha_\ell$
and $\beta_\ell \equiv c_\ell i_{\ell+1/2}(q R) \bar \beta_\ell$,
the functions $\bar \alpha_\ell$ and $\bar \beta_\ell$ are
determined by the solution of the following system of linear
equations:
\begin{subequations}
\label{alpa_beta_3d_eqs}
\begin{eqnarray}
\label{alpa_3d}
\bar \alpha_\ell + \bar \beta_\ell w_{\ell+1/2}(q R)
&=& v_{\ell+1/2}(q R),\label{BC_at_R}\\
\bar \alpha_\ell + \bar \beta_\ell w_{\ell+1/2}(\eta q R)
&=& w_{\ell+1/2}(\eta q R),\label{BC_at_R1}
\end{eqnarray}
\end{subequations}
where $v_{\ell+1/2}(z):= k_{\ell+1/2}(z)/i_{\ell+1/2}(z)$ and
$w_{\ell+1/2}(z):= [d k_{\ell+1/2}(z)/dz]/[d i_{\ell+1/2}(z)/dz]$.
This determines the Green's function $\bar G(\vect{r},q)$.

Before proceeding, we make three remarks:\\
\textbf(i) Since for $q R > 0$ one has
${\displaystyle \lim_{\eta \to \infty}} w_{\ell+1/2}(\eta q R) = 0$,
in the limit $\eta \to \infty$ one finds $\bar \alpha_\ell = 0$ and
Eq. (\ref{alpa_3d}) reduces to the corresponding BC in
Ref.~\cite{Golestanian_2005}, i.e., as expected the solution for the
unbounded case is recovered.\\
\textbf(ii) Equations (\ref{BC_at_R}) and (\ref{BC_at_R1}) will not
coincide in the limit $\eta \to 1$ because even in this limit a
source singularity is present at $r_s$ and is picked up by $G_s(q)$
[Eq.(\ref{Gs_3d_expand})].\\
\textbf(iii) The Laplace transform $\bar N(p)$ of the total number
of product particles in the system at time $t$ is
\begin{eqnarray}
\label{N_3d}
&&\bar N(p = D q^2) = 2\pi\int\limits_{R}^{R_1} \upd r r^2
\int\limits_{0}^{\pi} \upd \theta \sin\theta
P_0(\cos\theta) [-\bar a(p)] \bar G(q)\nonumber\\
&&= - 4 \pi \bar a(p) c_0(q) i_{1/2}(q R) \nonumber\\
&& \times \int\limits_{R}^{R_1} \upd r r^2
[\bar \alpha_0 i_{1/2}(q r) + (\bar \beta_0 - 1) k_{1/2}(q r)] \\
&&= D \bar a(p) /p \Rightarrow N(t) =
\int\limits_{0}^{t} \upd  t' A(t')\,, \nonumber
\end{eqnarray}
i.e., as expected $N(t)$ is given by the time integral of the
production rate, providing a welcome consistency check.

We now focus on the particular case in which the activity of the
catalytic site is time independent, i.e., $A(t)$ is constant:
$A(t) = H(t)/\tau_f$, where $H(t)$ is the Heaviside step
function and $\tau_f$ is the average production time for the creation
of a product molecule. Taking the Laplace transform of
Eq.~(\ref{def_force_3d}) and noting that $\cos \theta = P_1(\cos \theta)$,
one obtains
\begin{eqnarray}
\label{F_3d}
\bar F_z(p) &=& 2\pi b R^2 \bar a(p) \int\limits_{0}^{\pi}
\upd \theta \sin\theta P_1(\cos\theta) \bar G(q) \nonumber\\
&=& - F_0 \frac{R^2}{D} \bar \Phi\left(p \frac{R^2}{D};\eta\right)\\
&\Rightarrow& \dfrac{F_z\left(u=\dfrac{D t}{R^2}\right)}{F_0}
= -\left({\cal L}^{-1}[\bar \Phi (p;\eta)]\right)|_u\,,\nonumber
\end{eqnarray}
where
$F_0 := 2 b R/(\pi D \tau_f)$ and
\begin{eqnarray}
\label{Phi_def}
\bar \Phi (p;\eta) &=& \frac{i_{3/2}(\sqrt{p})}{\sqrt{p}}
\lbrace \bar\alpha_1(\sqrt{p};\eta) \,i_{3/2}(\sqrt{p})
\nonumber\\
&+&
[\bar\beta_1(\sqrt{p};\eta)-1] \,k_{3/2}(\sqrt{p})
\rbrace
\,.
\end{eqnarray}
At room temperature, using the order of magnitude estimates
from Ref.~ \cite{Golestanian_2005}, i.e., $b = k_B T$,
$R = 50~\mathrm{nm}$, $D = 5.5 \times 10^{-10} \mathrm{m}^2/\mathrm{s}$
(corresponding to a product-particle diameter
$\sigma = 0.8~\mathrm{nm}$ and a viscosity of the solvent of
$\mu = 1~\mathrm{mPa} \times \mathrm{s}$), and
$1/\tau_f = 25~\mathrm{kHz}$ leads to values of $F_0$ of the order of
$10^{-2}$ pN, which would correspond to a velocity
$v_0 \sim F_0/(\mu R)$ of the order of 100 nm/s, while, for comparison,
the weight $W$ of such a particle made out of latex
(density $\rho \simeq 10^3 \mathrm{kg/m}^3$, neglecting buoyancy) is
of the order of $10^{-6}$ pN, i.e., $F_0 \gg W$.

The complicated structure of the function $\bar \Phi (p;\eta)$
makes it rather laborious to carry out the inverse Laplace
transform, so that the full time dependence of the force $F_z$
cannot be derived easily. However, using the large $p$ expansion of
Eq. (\ref{Phi_def}) the short time behavior of $F_z$ can be obtained
from Eq. (\ref{F_3d}),
\begin{equation}
\label{Fshort_3d}
F_z(u\ll 1, \eta)/F_0 = \sqrt{\pi} \sqrt{u}\,,~u=D t/R^2\,.
\end{equation}
This dependence can be rationalized as being due to the diffusion-like
increase of the area on the particle around its catalytic site that is
covered by product molecules. On the other hand, the asymptotic value
$F_z^{(\infty)}(\eta):= {\displaystyle \lim_{u \to \infty}}
F_z(u;\eta)$, i.e., the ``steady state'' force, can be determined
using the inversion formula Eq.~(\ref{ILT}) by
noticing that $\bar \Phi (p;\eta)$ has a simple pole at $p = 0$,
which determines the asymptotic value as the residue of
$\bar \Phi (p;\eta)$ at $p=0$ \cite{Carslaw_book}. This yields
\begin{eqnarray}
\label{Fasymp_3d}
F_z^{(\infty)}(\eta)/F_0
&=& -\mathrm{Res}\,[\bar \Phi (p;\eta),p=0]\nonumber\\
\Rightarrow
F_z^{(\infty)}(\eta)/F_0
&=& \frac{\pi}{4} \frac{\eta^{3}+2}{\eta^{3}-1}\,,
\end{eqnarray}
which holds for any value of $\eta > 1$.
Together with Eqs. (\ref{F_3d}) and (\ref{Fasymp_3d}) the expansion
of Eq. (\ref{Phi_def}) for small $p$ renders the asymptotic temporal
approach of the force towards its steady-state value:
\begin{eqnarray}
\label{Flong_3d}
&&\frac{F_z(u\gg 1, \eta)}{F_z^{(\infty)}(\eta)} \simeq
1-e^{-u/\tau(\eta)}\,,\nonumber\\
&&\tau(\eta) = \frac{3}{10} \frac{\eta^5-1}{\eta^3-1}\,.
\end{eqnarray}
The corresponding relaxation time $T(\eta) = R^2/[D \tau(\eta)]$
is finite even in the case of extreme confinement, i.e.,
for $\eta \to 1$: $\tau(\eta \to 1) = 1/2$ and increases with
decreasing confinement (increasing $\eta$), as intuitively
expected.

\begin{figure}[!htb]
\onefigure[width=.9\linewidth]{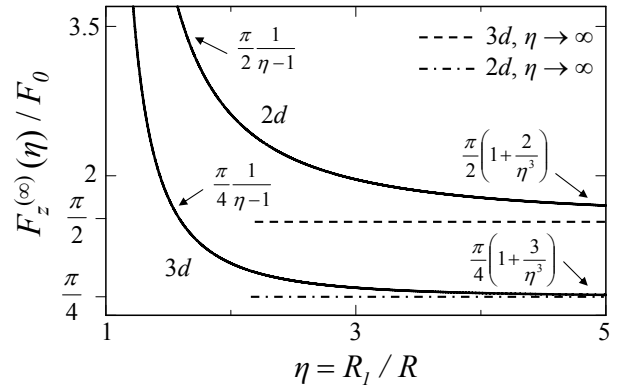}
\caption{
The steady-state force $F_z^{(\infty)}(\eta)/F_0$ as a function
of $\eta$.
The horizontal lines indicate the asymptotic values
$F_z^{(\infty)}(\eta\to\infty)/F_0 =$ $\pi/4$ in \textit{3d}
and $\pi/2$ in \textit{2d}, respectively.
}
\label{fig2}
\end{figure}

For systems of infinite size ($\eta \to \infty$) we recover the
result $F_z^{(\infty)}(\eta\to\infty)/F_0 = \pi/4$ in
Ref.~\cite{Golestanian_2005}. For all finite $\eta$ one has
$F_z^{(\infty)}(\eta) > F_z^{(\infty)}(\eta\to\infty)$.
Therefore the spatial confinement leads to an enhancement of the
diffusiophoretic propulsion, which is the main result of this
analysis. As shown in Fig.~\ref{fig2}, even at moderate values of
$\eta$ the effect of the confinement is strong, e.g., at
$\eta \simeq 2$ there is a ca. $25\%$ increase in the force
compared with the unconfined case. For $\eta \to 1$ the force
diverges; this is, of course, an artifact with stems from the
assumption that the product molecules are point-like. Actually,
there is a lower cut-off $R_1^c$, where $R_1^c - R$ is of the
order of the hard core diameter $\sigma$ of the product molecules,
below which this assumption breaks down and Eq. (\ref{Fasymp_3d})
is no longer valid. For $\sigma \ll R$, which is a reasonable
assumption, one has $\eta_c = 1 + \sigma/R \gtrsim 1$, and thus
the steep increase in the force near $\eta = 1$ is physically
relevant.

\noindent{\it The two-dimensional case}\\
\indent The analysis follows along essentially the same lines as
in \textit{3d}, so that we merely outline the steps involved
and present the final result. Before proceding, we note
that this case would correspond either to a very thin disk enclosed
by a circular wall, or to a macroscopicaly long cylinder with a
longitudinal line of catalyst on its surface enclosed by another
macroscopicaly long cylindrical wall. In \textit{2d}, the free space
Green's function for the Helmholtz operator is given by
\cite{Hassani_book}
\begin{equation}
\label{Gs_2d}
G_s(\vect{r},q;{\vect{r}}_s) =
-\frac{1}{2\pi} K_0(q|\vect{r}-{\vect{r}}_s|),
\end{equation}
and the regular part can be written as
\begin{eqnarray}
\label{g_2d}
g(r,\theta,q;{\vect{r}}_s,\eta)
&=& \sum\limits_{m \geq 0}
[\alpha_m(q;{\vect{r}}_s,\eta) I_m(q r) +
\\\nonumber
&+& \beta_m(q;{\vect{r}}_s,\eta) K_m(q r)] \cos (m\phi)\,.
\end{eqnarray}
Here, $K_m(z)$ and $I_m(z)$ are the modified Bessel functions
of integer index $m$ \cite{Abramowitz_book}, and the coefficients
$\alpha_m(q;{\vect{r}}_s,\eta)$ and $\beta_m(q;{\vect{r}}_s,\eta)$
will be fixed by the BCs. Note that in Eq. (\ref{g_2d}) there are
no terms proportional to $\sin (m\phi)$ because the solution has
to be invariant with respect to the transformation $\phi \mapsto -\phi$.

The addition theorem for the modified Bessel function $K_0$
\cite{Carslaw_book} together with the fact that the angle between
$\vect{r}$ and ${\vect{r}}_s$ is equal to $\pi-\phi$
(see the inset of Fig.~\ref{fig1}) allows one to represent the singular
part as the series
\begin{equation}
\label{Gs_2d_expand}
G_s(\vect{r},q;{\vect{r}}_s) = - \sum\limits_{m \geq 0}
c_m I_m(q r_<) K_{m}(q r_>) \cos (m\phi)
\end{equation}
where $c_m = (-1)^m (2-\delta_{m,0})/ (2\pi)$. Using Eqs. (\ref{g_2d}),
(\ref{Gs_2d_expand}), the BCs [Eqs. (\ref{BC_for_g})], and noting
that $r_s = R_+$, the coefficients
$\alpha_m \equiv c_m I_m(q R) \bar\alpha_m$ and
$\beta_m \equiv c_m I_m(q R) \bar\beta_m$ follow from the solution
of the system of linear equations
\begin{subequations}
\label{alpa_beta_2d_eqs}
\begin{eqnarray}
\bar \alpha_m + \bar \beta_m w_m(q R) &=& v_m(q R)\,,\\
\bar \alpha_m + \bar \beta_m w_m(\eta q R) &=& w_m(\eta q R)\,,
\end{eqnarray}
\end{subequations}
where $v_m(z):= K_m(z)/I_m(z)$ and
$w_m(z):= [d K_m(z)/dz]/[d I_m(z)/dz]$.
This fixes the Green's function $\bar G(\vect{r},q)$.

Similarly to the \textit{3d} case, in the limit $\eta \to \infty$
one recovers the result of a direct calculation for the unconfined
system, the limit $\eta \to 1$ still picks up the singularity at
${\vect{r}}_s$, and the total number of particles is given by the
time integral of the production rate.

Again focusing on the case of a time-independent source,
$A(t) = H(t)/\tau_f$, from the Laplace-transformed
Eq.~(\ref{def_force_2d}) one obtains
\begin{eqnarray}
\label{F_2d}
\bar F_z(p) &=& b R \bar a(p) \int\limits_{0}^{2 \pi} \upd \phi
(\cos\phi)\, \bar G(q) \nonumber\\
&=& - \frac{\pi}{2} F_0 \frac{R^2}{D}\,
\bar \Phi_1\left(p \frac{R^2}{D};\eta\right)\\
\Rightarrow &&\dfrac{F_z\left(u=\dfrac{D t}{R^2}\right)}{F_0}
= - \frac{\pi}{2}
\left({\cal L}^{-1}[\bar \Phi_1 (p;\eta)]\right)|_u\,,\nonumber
\end{eqnarray}
where
\begin{eqnarray}
\label{Phi1_def}
\bar \Phi_1 (p;\eta) &=& \frac{I_1(\sqrt{p})}{p} \times
\left[\bar\alpha_1(\sqrt{p};\eta) I_1(\sqrt{p})\right.\nonumber\\
&+& \left.(\bar\beta_1(\sqrt{p};\eta)-1) K_1(\sqrt{p}) \right]\,.
\end{eqnarray}

The short time behavior of $F_z$ is obtained from Eq. (\ref{F_2d})
by using the large $p$ expansion of Eq. (\ref{Phi1_def}). 
One finds the same expression [Eq. (\ref{Fshort_3d})] as 
in \textit{3d}, which simply reflects the increase over time in 
the ``support'' of the distribution of the product molecules on 
the surface of the particle. The ``steady state'' diffusiophoretic 
force
$F_z^{(\infty)}(\eta):=
{\displaystyle \lim_{u \to \infty}} F_z(u;\eta)$ follows from the
inversion formula in Eq.~(\ref{ILT}) and the fact that
$\bar \Phi_1 (p;\eta)$ has a simple pole at $p = 0$:
\begin{eqnarray}
\label{Fasymp_2d}
F_z^{(\infty)}(\eta)/F_0
&=& -\frac{\pi}{2} \mathrm{Res}\,[\bar \Phi (p;\eta),p=0]\nonumber\\
\Rightarrow
F_z^{(\infty)}(\eta)/F_0 &=& \frac{\pi}{2} \frac{\eta^{2}+1}{\eta^{2}-1}\,.
\end{eqnarray}
The long time behavior, i.e., the approach of $F_z$ to its asymptotic
value follows from Eqs. (\ref{F_2d}) and (\ref{Fasymp_2d}) by using
the small $p$ expansion in Eq. (\ref{Phi1_def}) leading to
\begin{eqnarray}
\label{Flong_2d}
&&\frac{F_z(u\gg 1, \eta)}{F_z^{(\infty)}(\eta)} \simeq 
1 - \,e^{-u/\tau(\eta)}\,,\nonumber\\
&& \tau(\eta) = \frac{1}{2}\frac{\eta^2 \ln (\eta)}{\eta^2-1}\,.
\end{eqnarray}
The relaxation time $T(\eta) = R^2/[D \tau(\eta)]$ stays finite 
even for extreme confinement because $\tau(\eta \to 1) = 1/4$, and 
it increases with decreasing confinement, but only as $\ln(\eta)$, 
i.e., much slower than the increase $\sim \eta^2$ in the \textit{3d} 
case (see Eq.(\ref{Flong_3d})). We note that occurrence of terms 
$\sim s \ln(s)$ or $\sim t/\ln(t)$ in the Laplace or the time domain, 
respectively, is generic for two-dimensional infinitely large 
reaction-diffusion systems \cite{log_correc} because the fractal 
dimension of Brownian paths coincides with the dimension of the 
embedding space. For confined systems, however, the 
logarithmic correction $\ln(t)$ to the relaxation time is replaced 
by $\ln(\eta)$, which is related to finite-size effects. 

Equation (\ref{Fasymp_2d}) implies that the value of the force
$F_z^{(\infty)}(\eta\to\infty)/F_0 = \pi/2$ in the unbounded case
is twice as large as the corresponding one in \textit{3d} (assuming
the same values for $R$, $D$, $b$, and $\tau_f$ in both cases). For any
$\eta$, the force $F_z^{(\infty)}(\eta)$ is stronger in \textit{2d}
than in \textit{3d}, which can be easily inferred from the ratio of
Eqs. (\ref{Fasymp_2d}) and (\ref{Fasymp_3d}), as shown in
Fig.~\ref{fig2}. Thus we conclude that reduced dimensionality enhances
the propulsion mechanisms. Such a two-dimensional system
can be realized, e.g.,  by using a disk-like particle floating on a
liquid-vapor interface and product particles that are mobile on the
interface but do not dissolve in the bulk liquid and do not evaporate
either.

As in the \textit{3d} case, with increasing confinement (i.e.,
decreasing $\eta$) the force increases steeply from the value
$F_z^{(\infty)}(\eta\to\infty)/F_0 = \pi/2$ and for $\eta \to 1$ it
diverges (up to a lower cut-off at $\eta_c \gtrsim 1 + \sigma/R$).

In summary, based on the model proposed in Ref. \cite{Golestanian_2005}
we have studied the effect of confinement on the diffusiophoretic
force exerted on a self-propelling particle in both three and two
spatial dimensions. The main findings (see Eqs. (\ref{Fasymp_3d}),
(\ref{Fasymp_2d}) and Fig.~\ref{fig2}) are:
\textbf{(1)} The presence of a confining wall for the reaction product
leads to a significant increase in the force acting on the particle,
even in the case of the simple concentric confining geometry
(see Fig.~\ref{fig1}).
\textbf{(2)} Reduced dimensionality alone (i.e., even without
confinement) is already sufficient to increase the efficiency of the
proposed propulsion mechanism, leading to a stronger force on the
particle. A possible way for testing such predictions would be to
use a colloidal particle with a catalyst grafted on its surface
and enclosed in shape-controllable vesicles filled with a solution
of reactants. By superposition the approach presented here can be
extended to the case of an arbitrary decoration of the particle
surface by catalytic sites as long as the reactions can be assumed
to occur independently, i.e., without competing for the reactants
in the common environment. This will allow one to find an optimal
decoration for providing stability against particle rotations,
which can otherwise spoil the unidirectional motion. Further
extensions may focus on new phenomena emerging from more
complicated geometries.

\acknowledgments
M.N.P. gratefully acknowledges very fruitful discussions with
A. Gambassi, L. Harnau, and M. Tasinkevych.

\end{document}